\documentclass[aps,prb,floatfix,twocolumn,superscriptaddress]{revtex4-2}
\usepackage{mathrsfs}
\usepackage{graphicx}
\usepackage[english]{babel}
\usepackage{amsmath}
\usepackage{amssymb}
\usepackage{xcolor}
\usepackage{cancel}
\usepackage[normalem]{ulem}

\usepackage{relsize}
\usepackage{CJK}
\usepackage{dsfont}
\usepackage{bm}

\newcommand{\me}{\mathrm{e}}
\newcommand{\mi}{\mathrm{i}}

\newcommand{\dif}{\mathrm{d}}

\newcommand{\bk}{\mathbf{k}}

\allowdisplaybreaks

\begin{document}

\title{Sj$\ddot{\text{o}}$qvist quantum geometric tensor of finite-temperature mixed states}

\author{Zheng Zhou}
\affiliation{School of Physics, Southeast University, Jiulonghu Campus, Nanjing 211189, China}

\author{Xu-Yang Hou}
\affiliation{School of Physics, Southeast University, Jiulonghu Campus, Nanjing 211189, China}

\author{Xin Wang}
\affiliation{School of Physics, Southeast University, Jiulonghu Campus, Nanjing 211189, China}

\author{Jia-Chen Tang}
\affiliation{School of Physics, Southeast University, Jiulonghu Campus, Nanjing 211189, China}

\author{Hao Guo}
\email{guohao.ph@seu.edu.cn}
\affiliation{School of Physics, Southeast University, Jiulonghu Campus, Nanjing 211189, China}
\affiliation{Hefei National Laboratory, University of Science and Technology of China, Hefei 230088, China}

\author{Chih-Chun Chien}
\email{cchien5@ucmerced.edu}
\affiliation{Department of physics, University of California, Merced, CA 95343, USA}
\affiliation{ITAMP, Center for Astrophysics, Harvard and Smithsonian, Cambridge, Massachusetts 02138, USA}

\begin{abstract}
The quantum geometric tensor (QGT) reveals local geometric  properties and associated topological information of quantum states.
Here a generalization of the QGT to mixed quantum states at finite temperatures based on the Sj$\ddot{\text{o}}$qvist distance is developed. The resulting Sj$\ddot{\text{o}}$qvist QGT is invariant under gauge transformations of individual spectrum levels of the density matrix. A Pythagorean-like relation connects the distances and gauge transformations, which clarifies the role of the parallel-transport condition. The real part of the QGT naturally decomposes into a sum of the Fisher-Rao metric and Fubini-Study metric, allowing a distinction between different contributions to the quantum distance. The  imaginary part of the QGT is proportional to a weighted summation of the Berry curvatures, which leads to a geometric phase for mixed states under certain conditions. We present three examples of different dimensions to illustrate the temperature dependence of the QGT and a discussion on possible implications.

\end{abstract}

\maketitle

\section{Introduction}
The quantum geometric tensor (QGT) characterizes the distance and local geometry of quantum states as a set of parameters changes~\cite{QGTCMP80,BRODY200119,QGT10,KOLODRUBETZ20171}. The QGT has played an increasingly important role in various fields of physics, including quantum statistics, quantum information, condensed matter physics, and atomic, molecular, and optical physics \cite{IG_Book,Bohm03,KOLODRUBETZ20171,QGTCMP80,cmp/1103904831,RevModPhys.82.1959,PhysRevResearch.3.L042018,PhysRevB.74.085308,PhysRevLett.72.3439,PhysRevB.103.014516, PhysRevB.103.205415, PhysRevB.102.155407}. Since quantum states are usually described by their amplitudes and phases, the QGT is in general a complex second-order tensor after gauge invariance with respect to the overall phase factor of the state has been taken care of. For pure states, the real and imaginary parts of the QGT are respectively the Fubini-Study metric \cite{EGUCHI1980213} and (proportional to) the Berry curvature \cite{Simon83,Berry84}. Thus, it can reveal local geometry \cite{PhysRevLett.99.100603,PhysRevLett.117.045303} and associate with globally topological features \cite{PhysRevB.90.165139,PhysRevA.92.063627,PhysRevB.105.045144,PhysRevLett.121.170401,PhysRevB.104.045103} by examining the resulting distances or integrals. It has also been discussed in non-Hermitian systems \cite{PhysRevA.99.042104}.
For pure states, the QGT has been found relevant to some physical observables via response functions or topological indicators \cite{PhysRevB.108.094508,PhysRevB.87.245103,PhysRevB.97.201117,PhysRevB.97.041108,PhysRevLett.121.020401,PhysRevB.97.195422,PhysRevLett.124.197002,PhysRevX.10.041041}. Those connections allow the pure-state QGT to be experimentally studied in many platforms, such as NV centers in diamonds~\cite{10.1093/nsr/nwz193}, superconducting qubits~\cite{PhysRevLett.122.210401}, exciton-photon polaritons~\cite{Gianfrate2020}, plasmons~\cite{Cuerda23}, and ultracold atoms~\cite{Yi23}. Furthermore, the pure-state QGT is behind many striking  phenomena, including open quantum systems with general Lindbladians \cite{PhysRevX.6.041031}, quantum phase transitions \cite{PhysRevLett.99.100603}, orbital magnetic susceptibility \cite{PhysRevB.91.214405,PhysRevB.94.134423}, and superfluidity on the Lieb-lattice \cite{PhysRevLett.117.045303}.

While most of the studies of the QGT focus on pure states, a generalization to mixed quantum states is necessary and inevitable since the latter is common in nature, including all finite-temperature systems in thermal equilibrium. Since two pure states differ by a phase factor should be considered physically equivalent, the  invariance of the QGT against local U(1)-transformations due to the phase factor of the wave-function guarantees the distinction between physically equivalent and inequivalent states. Thus, the real part of the gauge-invariant QGT measures the genuine quantum distance between physically inequivalent pure states. When generalizing to mixed states, a natural requirement is that the corresponding QGT is also invariant under suitable gauge transformations, so it can measure the distance between physically inequivalent mixed states.

One possible mixed-state QGT has been developed in Ref.~\cite{OurQGT23} based on the Uhlmann approach. Explicitly, the total and physical spaces of full-rank density matrices are characterized via purification and the Uhlmann parallel-transport condition \cite{Uhlmann86}. The phase factor arises from the polar decomposition of the amplitude of the density matrix, thereby introducing a U$(N)$ gauge transformation.  The U$(N)$-invariant QGT has a real part that reduces to the Bures metric and an imaginary part that vanishes for typical systems. By considering thermal states approaching the zero-temperature limit, the real-part of the U$(N)$-invariant QGT agrees with that of the pure-state QGT. In contrast, the imaginary part of the U$(N)$-invariant QGT is zero but that of the pure-state QGT is the Berry curvature, which is not necessarily zero.
The U$(N)$-invariant QGT is mathematically rigorous, but it may be quite restricted when applied to physical systems.

Recently, Sj\"oqvist introduced a distance between density matrices \cite{PhysRevResearch.2.013344}, which will be called the Sj\"oqvist distance. We will show that it is invariant under the U$^N(1)\equiv\underbrace{\text{U}(1)\times\cdots \times\text{U}(1)}_N$ gauge transformation for full-rank density matrices. Compared with the U$(N)$ invariance of the Uhlmann-based approach, the gauge-invariance condition of the Sj\"oqvist distance is more relaxed, which also makes it more experimentally feasible \cite{PhysRevResearch.2.013344}. Moreover, we will construct a U$^N(1)$- invariant QGT for mixed states, called the Sj\"oqvist QGT, based on the Sj\"oqvist distance. Its real part is a Riemannian metric that contains the contributions from the Fisher-Rao metric and the Fubini-Study metric. Interestingly, its imaginary part introduces a two-form that does not vanish. Moreover, in some situation, an integral of the imaginary part of the Sj\"oqvist QGT produces a geometric phase that belongs to the thermal Berry phase~\cite{OurTB}, which is different from the Uhlmann phase \cite{Uhlmann86,Uhlmann91} and the interferometric geometric phase  \cite{PhysRevLett.85.2845} of mixed states.

We will illustrate the U$^N(1)$-invariant QGT by solvable examples in 1D, 2D, and 3D. As will be explained later, the 1D case is special because the imaginary part of the QGT vanishes automatically. In general, a smooth peak at finite temperature appears in the real part of the QGT due to its asymptotic behavior in the low- and high- temperature limits. The 2D example shows the behavior of a geometric phase associated with the imaginary part of the QGT. The 3D $s$-wave Fermi superfluid gives an example of the smoothness of the QGT across a phase transition. The examples also elucidate the geometric structures under simple physical systems via the QGT.

The rest of the paper is organized as follows. In Section \ref{Sec.1}, a derivation of the Sj\"oqvist QGT is given based on the purification of density matrices, quantum distances between mixed states, and gauge transformations. The expressions of the real and imaginary parts of the Sj\"oqvist QGT are presented and analyzed.
In Section \ref{Sec.2}, three examples of different dimensions are provided to visualize the Sj\"oqvist QGT. Sec.~\ref{sec:implications} discusses experimental and theoretical implications of the QGT. Finally, Sec.~\ref{sec:conclusion} concludes the work. The Appendix gives some details and derivations.

\section{Basic formalism}\label{Sec.1}
\subsection{Purification of density matrix}
Before generalizing the QGT to mixed states, a key tool that provides a pure-state like description of the density matrix is briefly reviewed. A mixed quantum state is in general depicted by a Hermitian density matrix $\rho$ without explicitly information of any phase from the wavefunctions. To incorporate the effect of phases into mixed states, purification of density matrix has been frequently used in quantum information theory~\cite{Bengtsson_book}.
For a density matrix $\rho$, its purification (or amplitude) is defined as $W=\sqrt{\rho}U$ or conversely $\rho=WW^\dag$, where $U$ is an arbitrary unitary matrix often referred to as the phase factor. The relation is also known as the polar decomposition of $W$, which is uniquely determined if $\rho$ is full rank. Our discussion will focus on full-rank density matrices, which cover systems in thermal equilibrium.

Purification of $\rho$ has a U$(N)$-degrees of freedom since $\rho$ is invariant under a transformation $W\rightarrow W'=W\mathcal{U}$ with $\mathcal{U}\in \text{U}(N)$. Therefore, purification allows phase effects of mixed states to be introduced like those of pure states.
By diagonalizing the density matrix as $\rho=\sum_{n=0}^{N-1}\lambda_n|n\rangle\langle n|$, purification is expressed by $W=\sum_{n=0}^{N-1}\sqrt{\lambda_n}|n\rangle\langle n|U$, where $N=\text{rank}(\rho)$. $W$ is isomorphic to
$|W\rangle=\sum_n\sqrt{\lambda_n}|n\rangle\otimes U^T|n\rangle$, which is known as the purified state. Furthermore, one can introduce the inner product between two purified states via the Hilbert-Schmidt product $\langle W_1|W_2\rangle=\text{Tr}(W^\dagger_1 W_2)$. We will set $\hbar=k_B=1$ and use the convention of Einstein summation over repeated indices in the subsequent discussions.

\subsection{Quantum distances between mixed states}
Through purification, several types of distances between density matrices have been developed. Assuming $W$ (or equivalently, $\rho$,) continuously depends on a set of real-valued parameters $\mathbf{R}=(R^1,R^2,\cdots, R^k)^T$, the ``raw'' distance between $W(\mathbf{R}+\dif\mathbf{R})$ and $W(\mathbf{R})$ is introduced via the Hilbert-Schmidt product:
\begin{align}\label{Sdis3b}
&\dif^2(W(\mathbf{R}+\dif \mathbf{R}),W(\mathbf{R}))=\Big||W(\mathbf{R}+\dif \mathbf{R})\rangle-|W(\mathbf{R})\rangle\Big|^2\notag\\
=&\langle \partial_\mu W(\mathbf{R})|\partial_\nu W(\mathbf{R})\rangle \dif R^\mu \dif R^\nu.
\end{align}
We refer to $g_{\mu\nu}\equiv\langle \partial_\mu W|\partial_\nu W\rangle$ as the ``raw'' metric.
It is evident that neither this distance nor $g_{\mu\nu}$ is invariant under a local gauge transformation $W'(\mathbf{R})=W(\mathbf{R})\mathcal{U}(\mathbf{R})$ with $\mathcal{U}(\mathbf{R})\in \text{U}(N)$.
Consequently, Eq.~(\ref{Sdis3b}) does not measure the distance between physically inequivalent mixed states. Proper corrections must be imposed to eliminate the extra gauge redundancy.

The gauge-invariance problem has been encountered in pure states, and a proper solution has been developed to establish a U$(1)$ gauge-invariant metric via the quantum geometric tensor \cite{QGT10}. For mixed states, there have also been several methods to address this challenge. One approach is to take the infimum of the raw distance, leading to the Bures distance between different density matrices \cite{Uhlmann86}:
\begin{align}\label{Bdis}
&\dif^2_\text{B}(\rho(\mathbf{R}+\dif \mathbf{R}),\rho(\mathbf{R}))\notag\\=&\inf_{\mathcal{U}\in \text{U}(N)} \big||W(\mathbf{R}+\dif \mathbf{R})\rangle-|W(\mathbf{R})\rangle\big|^2.
\end{align}
The infimum is taken with respect to all possible phase factors $\mathcal{U}\in \text{U}(N)$.
Another method follows the standard procedure used for pure states and introduces a U($N$) gauge-invariant metric by utilizing the formalism of the Uhlmann bundle \cite{OurQGT23}.
The distance derived from this method is named the Uhlmann distance.
Interestingly, the Uhlmann distance reduces to the Bures distance when restricted on the base manifold of the Uhlmann bundle.

Yet another distance of mixed states has been proposed by Sj$\ddot{\text{o}}$qvist~\cite{PhysRevResearch.2.013344}, which will be generalized to a gauge-invariant QGT in our following discussions.

\subsection{The Sj$\ddot{\text{o}}$qvist distance}
We first briefly review the original construction of the Sj$\ddot{\text{o}}$qvist distance before generalizing it.
A smooth path $\mathbf{R}(t)=(R^1(t),R^2(t),\cdots, R^k(t))^T$ in the parameter manifold induces an evolving mixed state $\rho(t)\equiv \rho(\mathbf{R}(t))$. In this work, we focus on full-rank density matrices.
With the instantaneous eigenstates, the diagonal form is $\rho(t)=\sum_{n=0}^{N-1}\lambda_n(t)|n(t)\rangle\langle n(t)|$. Following Ref.~\cite{PhysRevResearch.2.013344}, one may introduce $N$ spectral rays $\{\me^{\mi\theta_n(t)}|n(t)\rangle\}$ ($n=0,1,\cdots$ $N-1$) along the path $\mathbf{R}(t)$ and
let $\mathcal{B}(t)=\{\sqrt{\lambda_n(t)}\{\me^{\mi\theta_n(t)}|n(t)\rangle\}_{n=0}^{N-1}$ be the spectral decomposition along the path.
The Sj$\ddot{\text{o}}$qvist distance is defined as the minimum distance between $\mathcal{B}(t)$ and $\mathcal{B}(t+\dif t)$:
\begin{align}\label{Sdis}
\dif^2_\text{S}(t+\dif t,t)&=\inf_{\theta_n}\sum_{n=0}^{N-1}\big|\sqrt{\lambda_n(t+\dif t)}\me^{\mi\theta_n(t+\dif t)}|n(t+\dif t)\rangle\notag\\
&-\sqrt{\lambda_n(t)}\me^{\mi\theta_n(t)}|n(t)\rangle\big|^2 \notag \\
&=2-2\sup\sum_n\sqrt{\lambda_n\lambda_n(t)}|\langle n|n(t)\rangle|\cos \phi_n(t).
\end{align}
The infimum is taken among all possible sets of spectral phases $\{\theta_n(t),\theta_n(t+\dif t)\}$.
Here $\cos \phi_n(t)=\dot{\theta}_n(t)\dif t+\arg\left[1+\langle n(t)|\dot{n}(t)\rangle \dif t\right]+O(\dif t^2)$. Thus, the infimum is obtained if
\begin{align}\label{Sdispc}
\mi\dot{\theta}_n(t)+\langle n(t)|\dot{n}(t)\rangle=0, \quad \text{for} \quad n=0,\cdots, N-1.
\end{align}
This is precisely the parallel-transport condition associated with each individual spectral level. Some details are in Appendix \ref{appa}.

Via purification, the Sj$\ddot{\text{o}}$qvist distance can be derived in a more instructive manner. By encoding a specific set of phase factors into the unitary operator $U(t)=\sum_n\me^{\mi\theta_n(t)}|n(t)\rangle\langle n(0)|$, $\rho(t)$ can be purified by the amplitude  \begin{align}W(t)&=\sum_n\sqrt{\lambda_n(t)}|n(t)\rangle\langle n(t)|U(t)\notag\\&=\sum_n\sqrt{\lambda_n(t)}|n(t)\rangle\langle n(0)|\me^{\mi\theta_n(t)}, \end{align}
 which corresponds to the purified state
 \begin{align}\label{Wt}
|W(t)\rangle=\sum_n\sqrt{\lambda_n(t)}\me^{\mi\theta_n(t)}|n(t)\rangle\otimes|n(0)\rangle.
 \end{align}
Similar to the Bures distance given by Eq.~(\ref{Bdis}), we introduce a $\text{U}^N(1)$-invariant distance:
\begin{align}\label{Sdis3}
\dif^2_{\text{U}^N(1)}(t+\dif t, t)&=\inf_{\theta_n}\big||W(t+\dif t)\rangle-|W(t)\rangle\big|^2 \notag \\
&=\inf_{\mathcal{U}\in \text{U}^N(1)}\big||W(t+\dif t)\rangle-|W(t)\rangle\big|^2.
\end{align}
Here the infimum is obtained with respect to the gauge transformation $W'(t)=W(t)\mathcal{U}(t)$, where $\mathcal{U}(t)=\sum_n\me^{\mi\chi_n}|n(0)\rangle\langle n(0)|$.
We note that the second spectral state $|n(0)\rangle$ in $|W(t)\rangle$ is independent of $t$ and gives no contribution to the local distance. One may observe that $\dif^2_{\text{U}^N(1)}$ is indeed equal to $\dif^2_\text{S}$, the Sj$\ddot{\text{o}}$qvist distance, by comparing Eq.~(\ref{Sdis3}) with Eq.~(\ref{Sdis}).

\subsection{Decomposition of distances}
The Sj$\ddot{\text{o}}$qvist distance can also be derived from a geometric point of view.
Using the compact notations $|n\rangle\equiv|n(\mathbf{R}(t))\rangle$ and $|n_0\rangle\equiv|n(\mathbf{R}(0))\rangle$ and applying
\begin{align}\label{Wep}
|\partial_\mu W\rangle=\sum_n\big[&\partial_\mu\sqrt{\lambda_n}\me^{\mi\theta_n}|n\rangle+\sqrt{\lambda_n}\me^{\mi\theta_n}|\partial_\mu n\rangle\notag\\+&\mi\sqrt{\lambda_n}\me^{\mi\theta_n}\partial_\mu\theta_n|n\rangle\big]\otimes |n_0\rangle
\end{align}
and the identity $\sum_n \sqrt{\lambda_n}\partial_\mu \sqrt{\lambda_n}=\frac{1}{2}\partial_\mu \sum_n\lambda_n=0$, it can be shown that the raw distance (\ref{Sdis3b}) is
\begin{align}\label{Sdis3c}
&\dif^2(W+\dif W,W)=\sum_n\big[\partial_\mu \sqrt{\lambda_n}\partial_\nu \sqrt{\lambda_n}+\lambda_n(\langle\partial_\mu n|\partial_\nu n\rangle\notag\\&+\partial_\mu \theta_n\partial_\nu\theta_n -\mi\omega_{n\mu}\partial_\nu\theta_n-\mi\omega_{n\nu}\partial_\mu\theta_n) \big]\dif R^\mu \dif R^\nu.
\end{align}
Here $\omega_{n\mu}=\langle n|\partial_\mu n\rangle=-\langle \partial_\mu n| n\rangle$ is the component form of the Berry connection $\omega_n=\langle n|\dif|n\rangle$ of the $n$th spectral level. In terms of differential forms, it can also be equivalently expressed as
 \begin{align}\label{Sdis3d}
&\dif^2(W+\dif W,W)=\sum_n\big\{(\dif \sqrt{\lambda_n})^2\notag\\&+\lambda_n\big[\langle\dif n|\dif n\rangle+(\dif\theta_n)^2-2\mi\omega_n\dif\theta_n\big]\big\}.
\end{align}
Accordingly, the minimizing condition (\ref{Sdispc}) is cast into the form
 \begin{align}\label{pxc4}
 \partial_\mu\theta_n-\mi\omega_{n\mu}=0,\quad \text{or} \quad \dif\theta_n-\mi\omega_n=0.
\end{align}
Under the minimum condition, the Sj$\ddot{\text{o}}$qvist distance is
\begin{align}\label{Sdis4}
\dif^2_\text{S}(\rho+\dif \rho,\rho)=&\sum_n\left\{(\dif \sqrt{\lambda_n})^2+\lambda_n\left[\langle\dif n|\dif n\rangle+(\omega_n)^2\right]\right\}\notag\\
=&\dif^2_\text{FR}+\sum_n\lambda_n\dif^2_{\text{FS}n}.
\end{align}
Here
\begin{align}\label{dFR}
\dif^2_\text{FR}(\rho+\dif \rho,\rho)=\sum_n(\dif \sqrt{\lambda_n})^2=\sum_n\frac{(\dif\lambda_n)^2}{4\lambda_n}
\end{align}
is the Fisher-Rao distance \cite{{Bengtsson_book}} representing the contribution from the thermal distribution, and
\begin{align}\label{dFS}
\dif^2_{\text{FS}n}=\langle\dif n|\dif n\rangle+(\omega_n)^2=\langle\dif n|(1-|n\rangle\langle n|)\dif n\rangle
\end{align}
is the Fubini-Study distance \cite{Uhlmann95} of the $n$th spectral level.

Comparing Eqs.~(\ref{Sdis3d}) and (\ref{Sdis4}), we come to a decomposition of the raw distance:
\begin{align}\label{Sdis3e}
\dif^2(W+\dif W,W)=\dif^2_\text{S}(\rho+\dif\rho,\rho)+\sum_n\lambda_n(\dif\theta_n-\mi\omega_n)^2.
\end{align}
If each spectral level undergoes parallel transport according to Eq.~(\ref{Sdispc}), or equivalently Eq.~(\ref{pxc4}), no contribution from the phase factor of each spectral level adds to the total distance since $\me^{\mi\theta_n(t+\dif t)}|n(t+\dif t)\rangle$ is kept in phase with $\me^{\mi\theta_n(t)}|n(t)\rangle$ in this case. Following parallel transport, the raw distance thus reduces to the Sj$\ddot{\text{o}}$qvist distance for physically inequivalent mixed states.
Interestingly, a similar decomposition that connects the raw distance and the U$(N)$-invariant Bures distance has been discussed in Ref.~\cite{OurQGT23}.

\subsection{The Sj$\ddot{\text{o}}$qvist QGT}
Eq.~(\ref{Sdis4}) leads to the Sj$\ddot{\text{o}}$qvist metric
\begin{align}\label{Sm0}
g^\text{S}_{\mu\nu}=&\sum_n\Big[\frac{\partial_\mu\lambda_n\partial_\nu\lambda_n}{4\lambda_n}+\lambda_n\langle\partial_\mu n|\partial_\nu n\rangle\notag\\-&\lambda_n\langle\partial_\mu n|n\rangle\langle n|\partial_\nu n\rangle)\Big],
\end{align}
where only the term $\langle\partial_\mu n|\partial_\nu n\rangle$ is complex-valued. By symmetrizing and antisymmetrizing the indices $\mu$ and $\nu$, its real and imaginary parts can be obtained. Moreover, using Eqs.~(\ref{dFR}) and (\ref{dFS}), we can decompose the Sj$\ddot{\text{o}}$qvist metric into
\begin{align}\label{Sm1}
g^\text{S}_{\mu\nu}=g^\text{FR}_{\mu\nu}+g^\text{FS}_{\mu\nu}-\mi\Omega_{\mu\nu},
\end{align}
where
\begin{align}
g^\text{FR}_{\mu\nu}=\sum_n\frac{\partial_\mu\lambda_n\partial_\nu\lambda_n}{4\lambda_n}
\end{align}
is the Fisher-Rao metric,
 \begin{align}
 g^\text{FS}_{\mu\nu}&=\sum_n\lambda_n g^{\text{FS}}_{n\mu\nu}=\sum_n\lambda_n(\text{Re}\langle\partial_{\mu}n| \partial_\nu n\rangle+\omega_{n\mu}\omega_{n\nu})
\end{align}
is the weighted summation of the Fubini-Study metrics from all spectral components, and
\begin{align}\label{iofSm}
\Omega_{\mu\nu}=&\frac{\mi}{2}\sum_n\lambda_n\left(\langle\partial_\mu n|\partial_\nu n\rangle-\langle\partial_\nu n|\partial_\mu n\rangle\right)\notag\\\equiv &\frac{1}{2}\sum_n\lambda_nF_{n\mu\nu}
\end{align}
is half of a weighted summation of all Berry curvatures
\begin{align}\label{FSt}
F_{n\mu\nu}\equiv\mi\partial_\mu\omega_{n\nu}-\mi\partial_\nu\omega_{n\mu}.
\end{align}
We note that $g^\text{FR}_{\mu\nu}$ and $g^\text{FS}_{\mu\nu}$ are both symmetric tensors and belong to the real part of the Sj$\ddot{\text{o}}$qvist metric. Accordingly, they both contribute to the Sj$\ddot{\text{o}}$qvist distance.
Meanwhile, $\Omega_{\mu\nu}$ is an anti-symmetric tensor and is the negative imaginary part of $g^\text{S}_{\mu\nu}$. It makes no contribution to the Sj$\ddot{\text{o}}$qvist distance.

The gauge invariance of the Sj$\ddot{\text{o}}$qvist metric can also be explicitly verified by modifying the raw metric $g_{\mu\nu}$ from Eq.~(\ref{Sdis3b}) and provides another construction inspired by the formalism of the U(1)-invariant QGT of pure states \cite{QGT10}. The details are summarized in Appendix~\ref{appb}.
Since the Sj$\ddot{\text{o}}$qvist metric contains both local geometric information via $g^\text{FR}_{\mu\nu}$ and $g^\text{FS}_{\mu\nu}$ and possibly topological information via $\Omega_{\mu\nu}$, we refer to the Sj$\ddot{\text{o}}$qvist metric $g^\text{S}_{\mu\nu}$ as the U$^N$(1)-invariant QGT for mixed states, or simply the Sj$\ddot{\text{o}}$qvist QGT, in the following.

\subsection{$\Omega_{\mu\nu}$ and its integrals}
Although the imaginary part of the Sj$\ddot{\text{o}}$qvist QGT, $\Omega_{\mu\nu}$, does not contribute to the Sj$\ddot{\text{o}}$qvist distance, its surface integral may result in a geometric phase.
As we have pointed out before, $\Omega_{\mu\nu}$ is the weighted summation of the Berry curvatures $F_{n\mu\nu}$, $n=0,1,\cdots, N-1$. Since the Berry curvature is a field strength tensor, an interesting question is whether $\Omega_{\mu\nu}$ from the Sj$\ddot{\text{o}}$qvist QGT is also a valid field strength tensor of some gauge field. For the pure-state case, the answer is affirmative \cite{QGT10}.
Although Eq.~(\ref{FSt}) shows that $F_{n\mu\nu}$  can be expressed as a field strength of the gauge field $\omega_n$, it does not imply
$\Omega_{\mu\nu}=\mi\partial_\mu(\sum_n\lambda_n\omega_{n\nu})-\mi\partial_\nu(\sum_n\lambda_n\omega_{n\mu})$
since the derivatives of $\lambda_n$ do not necessarily vanish. The next attempt is instead to find a gauge field $\mathcal{A}_\mu$ such that
$\Omega_{\mu\nu}=\mi\partial_\mu \mathcal{A}_\nu-\mi\partial_\nu \mathcal{A}_\mu$.
In general, both $\lambda_n$ and $\omega_{n\mu}$ are unknown functions of $\mathbf{R}$. Hence, an explicit solution to the aforementioned equation may not exist. Nevertheless, since $\Omega_{\mu\nu}$ is U$^N(1)$ gauge-invariant,  it may reveal some global features of the system. Moreover, its surface integral should also be gauge-invariant.

To quantify what $\Omega_{\mu\nu}$ entails, we introduce the 2-form $\Omega=\frac{1}{2}\Omega_{\mu\nu}\dif R^\mu\wedge \dif R^\nu=\frac{1}{2}\sum_n\lambda_n F_n$ with $F_n=\frac{1}{2}F_{n\mu\nu}\dif R^\mu\wedge \dif R^\nu$. Similar to the real part of the QGT, $\Omega$ also possesses some interesting features. For example, if the parameter space forms a 2D manifold, $\Omega$ must be closed because $\dif \Omega$ is a 3-form, which necessarily vanishes on a 2D manifold. 
When the dimension of the parameter space is greater than two, determining whether
$\Omega$ is a closed is hard because a solution of the gauge field $\mathcal{A}_\mu$ may not be easily determined either. 
Furthermore, since $\Omega$ is also non-degenerate and skew-symmetric, it is a symplectic form \cite{Nakahara} in the 2D case.
When the system evolves along a loop $C$ in a 2D parameter space, the integral of $\Omega_{\mu\nu}$ over a surface $S$ enclosed by $C$ is
\begin{align}\label{aofSm}
\theta_g(C)=\int_S\Omega=\frac{1}{2}\sum_n\int_S\lambda_nF_{n}.
\end{align}
If all $\lambda_n$ are constant over the area $S$,
\begin{align}\label{aofSm3}
\theta_g(C)=\frac{1}{2}\sum_n\lambda_n\int_SF_{n}=\frac{1}{2}\sum_n\lambda_n\theta_{\text{B}n}(C).
\end{align}
Here $\theta_{\text{B}n}(C)$ is the Berry phase associated with the $n$th spectral level. In this particular case, $\theta_g(C)$ represents half the weighted summation of all Berry phases, indicating its nature as a geometric phase. Following this clue, we employ $\theta_g(C)$ from Eq.~(\ref{aofSm}) to search for internal geometric information of the system.
A careful comparison shows that $\theta_g(C)$ is different from two geometric phases of mixed states commonly found in the literature, the Uhlmann phase \cite{Uhlmann86} and the interferometric geometric phase \cite{PhysRevLett.85.2845}.
Interestingly, the special case of Eq.~\eqref{aofSm3} belongs to the thermal Berry phase, whose general definition has been introduced in Ref.~\cite{OurTB}. In general situations, however, $\theta_g(C)$ is no longer (half of) the weighted summation of the Berry phases.

We remark that systems with only a single parameter in parallel transport should be viewed as a special case. Let $R_1$ be the only variable in parallel transport, the QGT then has just one component $g^\text{S}_{11}$. Moreover, the imaginary part must vanish because the indices of the latter is anti-symmetric. This is consistent with the observation that the Berry curvature as a 2-form vanishes on a 1D manifold. However, the Berry phase can still be evaluated in the single-parameter case from the holonomy although it is not directly related to the QGT.

\section{Examples}\label{Sec.2}

\begin{figure}[t]
\centering
\includegraphics[width=3.2in]{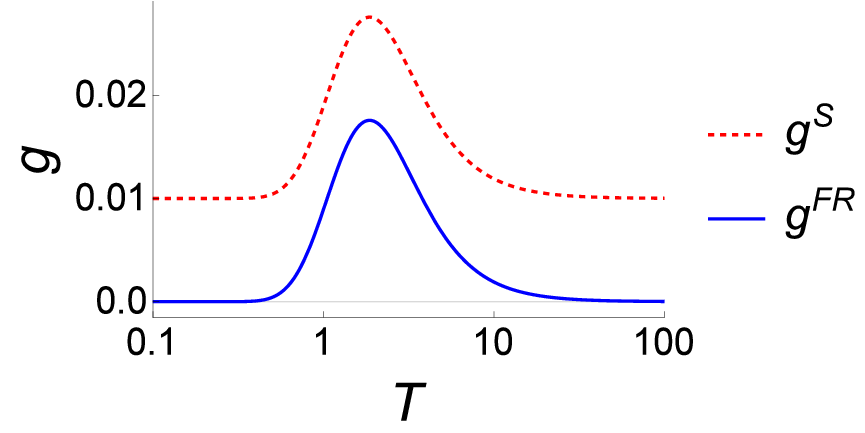}
\includegraphics[width=3.2in]{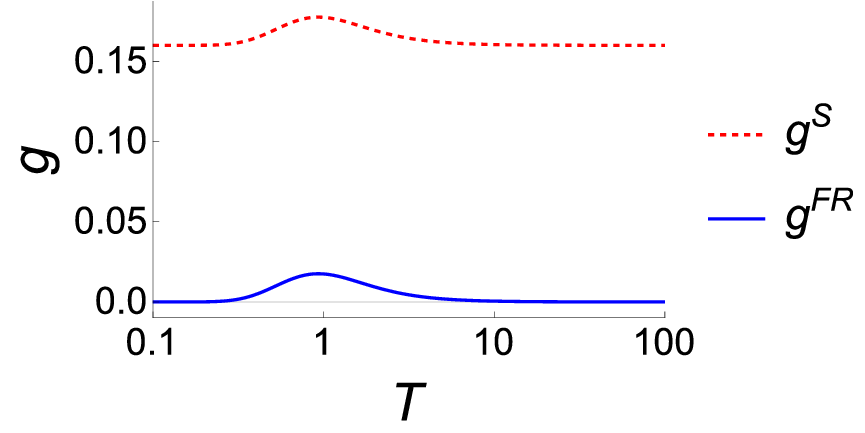}
\caption{Top panel: $g^\text{S}$ (dashed line) and $g^\text{FR}$ (solid line) of the 1D SSH model as functions of $T$ in units of $J_2$ with $k=\pi/2$ and $r=J_1/J_2=2.0$. Bottom panel: Same plot for $r=0.5$.
}
\label{Fig1}
\end{figure}

\subsection{1D Su-Schrieffer-Heeger model}
Our first example to demonstrate properties of the Sj$\ddot{\text{o}}$qvist QGT at finite temperatures is the one-dimensional (1D) Su-Schrieffer-Heeger (SSH) model \cite{SSH}, which is described by the following Hamiltonian with periodic boundary condition:
$\hat{H}=\sum^{L}_{i=1}(J_{1} a^{\dag}_{i}b_{i}+J_{2} a^{\dag}_{i}b_{i-1}+\text{H.c.} )$.
Here the alternating hopping coefficients $J_{1,2}$ are both positive. In momentum space, the SSH model can be written as $\hat{H}=\mathlarger{\int}_{0}^{2\pi} \frac{\mathrm{d} k}{2 \pi} \Psi_{k}^{\dagger} H(k) \Psi_{k}$, where $\Psi_k=(b_k,-a_k)^T$ is a Nambu spinor, and $H(k)= \mathbf{d}(k) \cdot \boldsymbol{\sigma}$ with $\mathbf{d}(k) =\left(-J_{1}-J_{2} \cos k, J_{2} \sin k, 0\right)^{T} $. The SSH model exhibits different topological properties between the regimes with $J_1/J_2>1$ and $J_1/J_2<1$. Hence, we introduce the dimensionless parameter $r=J_1/J_2$.

The two eigenvalues and their associated energy levels are respectively given by
\begin{align}\label{2b}
E_{\pm}\equiv\pm \tilde{R} J_2=\pm J_2\sqrt{1+r^2+2r\cos k},
\end{align}
and
\begin{equation}\label{3b}
\left|u_{\pm}\right\rangle=\frac{1}{\sqrt{2}\tilde{R}}\left(\begin{array}{c}
 \tilde{R}\\
\mp(r+\me^{-\mi k})
\end{array}\right).
\end{equation}
At temperature $T$ with $\beta=\frac{1}{T}$, the thermal equilibrium state is represented by $\rho=\frac{1}{2}\left[1-\tanh(\beta J_2\tilde{R})\hat{\mathbf{R}}_k\cdot\boldsymbol{\sigma}\right]$, whose eigenvalues are $\lambda_{\pm}=\frac{1}{2}[1\mp\tanh(\beta J_2\tilde{R})]$. In this model, we choose the momentum $k$ as the parameter to calculate the Sj$\ddot{\text{o}}$qvist QGT. As mentioned before, for the 1D SSH model with only a single parameter $k$ in parallel transport, the QGT has only one component $g^\text{S}_{kk}$, and its imaginary part vanishes.

Using Eq.~(\ref{Sm0}), a straightforward calculation shows $g^\text{S}_{kk}=g^\text{FR}_{kk}+g^\text{FS}_{kk}$, where
\begin{align}
g^\text{FR}_{kk}&=\text{sech}^2(\beta J_2\tilde{R})\beta^{2} J_2^{2}\frac{r^{2}\sin^{2} k}{4\tilde{R}^{2}},\notag\\
g^\text{FS}_{kk}&=\frac{(r\cos k+1)^{2}}{4\tilde{R}^4}.
\end{align}
Interestingly, the contribution from the Fubini-Study metric is independent of temperature. The expression shows that no significant difference appears when the regime changes from $r<1$ to $r>1$. This is in contrast to a change of the Berry phase by $\pi$ from the Berry holonomy since the real part of the Sj$\ddot{\text{o}}$qvist QGT only reveals local properties, unlike the topological indicator that reflects global properties. Moreover, the vanishing imaginary part of the single-parameter case like the 1D SSH model limits its information of topology as a special case. 

We emphasize that the Sj$\ddot{\text{o}}$qvist QGT is invariant under the U$^N(1)$ gauge transformation from a local phase transformation of the $N$ spectral rays of states. However, it is not invariant under a gauge transformation of the Bloch Hamiltonian with $H(k)\rightarrow U_k H(k)U^\dag_k$ and $\Psi_k\rightarrow U_k\Psi_k$, where the introduction of the unitary operator $U_k$ mixes the contributions within $\Psi_k$ but leaves the total Hamiltonian $\hat{H}$ invariant. Under the aforementioned transformation, it can be shown that $g^\text{FR}_{kk}$ remains invariant, but $g^\text{FS}_{kk}$ changes according to
\begin{align}
g^\text{FS}_{kk}\rightarrow g^\text{FS}_{kk}+\sum_{i=+,-}\lambda_{i}&\Big(2\text{Re}\langle \partial_{k} u_{i}|U^{\dagger}_{k}\partial_{k} U_{k}|u_{i}\rangle\notag\\&+\langle u_{i}|\partial_{k}U^{\dagger}_{k}\partial_{k} U_{k}|u_{i}\rangle\Big).
\end{align}

Figure~\ref{Fig1} shows the quantitative behavior of the QGT of the SSH model. The top and bottom panels show $g^\text{S}_{kk}$ and $g^\text{FR}_{kk}$ as functions of $T$ with $r=2.0$ (top) and 0.5 (bottom).
 Since $g^\text{FS}_{kk}$ remains a constant in this case, $g^\text{FR}_{kk}$ shows identical behavior to $g^\text{S}_{kk}$ with a vertical shift.
They all exhibit a peak at finite temperature that will be explained here.
As $T\rightarrow 0$, $\lambda_0\rightarrow 1$ and $\lambda_{n>0}\rightarrow 0$, thus $g^\text{FR}_{kk}\rightarrow0$ and the Sj$\ddot{\text{o}}$qvist QGT approaches the Fubini-Study metric of the ground state. As $T\rightarrow +\infty$, $\rho\rightarrow \frac{1}{N}1_N$, all density matrices converge to the $N\times N$ identity matrix. Hence, $g^\text{FR}_{kk}$ also approaches $0$ in this limit. Since $g^\text{FR}_{kk}(T\rightarrow 0)=g^\text{FR}_{kk}(T\rightarrow \infty)=0$, there must be at least a maximum at finite temperature.
Moreover, the bottom panel confirms that the Sj$\ddot{\text{o}}$qvist QGT is insensitive to the topological phase transition point at $r=1.0$.
where the Berry phase jumps. 
The change of $r$ only changes the vertical shift between $g^\text{FR}_{kk}$ and $g^\text{S}_{kk}$. We caution that a mixed-state generalization of the geometric phase, known as the Uhlmann phase, exhibits a finite-temperature transition in the topological regime of the SSH model~\cite{Viyuela14}.
In contrast, the QGT only depicts local geometric features of the quantum states and does not reveal the finite-temperature geometric phase transition of the SSH model. Therefore, mixed-state QGT and topological or geometric indicators may complement each other to unveil interesting physics of quantum systems at finite temperatures.

\begin{figure}[t]
\centering
\includegraphics[width=3.0in]{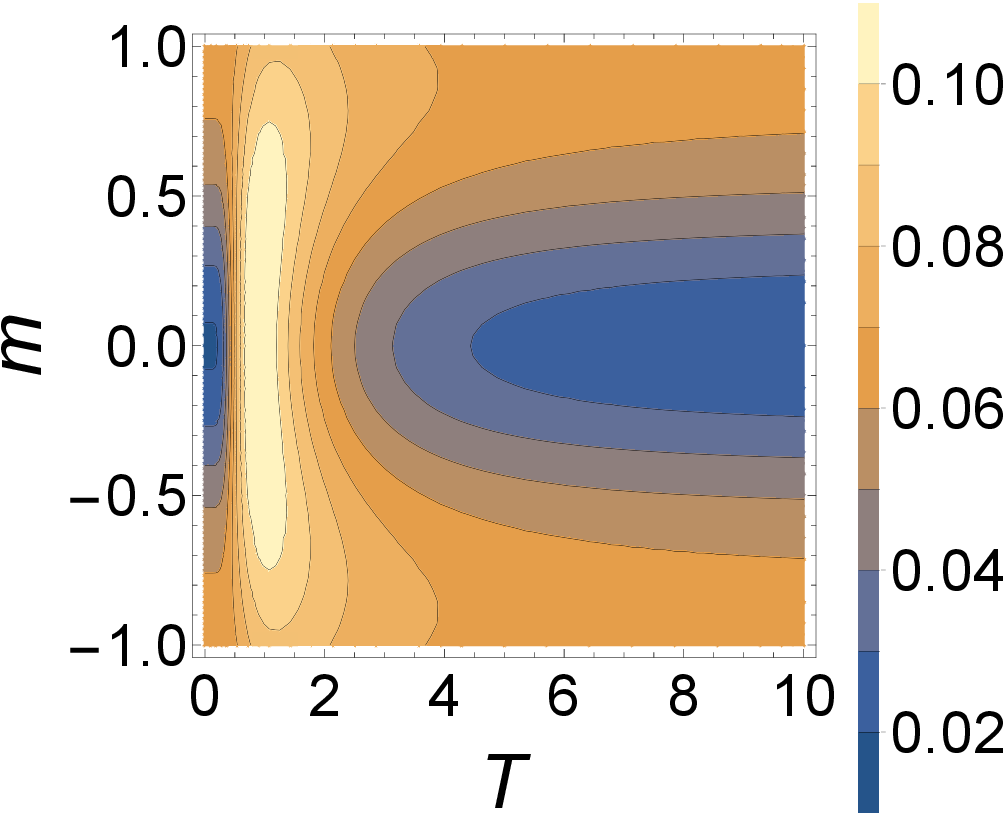}
\caption{Contour plot of $g^\text{S}_{11}$ of the 2D model, where $k_x=1.0$ and $k_y=0.3$. }
\label{Fig2a}
\end{figure}

\subsection{2D Dirac-fermion model}
Our second example is a 2D Dirac-fermion system with the Hamiltonian $H(\bk)= \mathbf{d}(\bk) \cdot \boldsymbol{\sigma}$, where $d_1=k_x$, $d_2=k_y$, and $d_3=m$. The eigen-energies and their eigenstates are given by $E_{\pm}(\bk)=\pm d(\bk)=\pm\sqrt{k^2+m^2}$ and
\begin{align}\label{EL}
|u_\pm\rangle=\frac{1}{\sqrt{2d(d\pm d_3)}}\left(\begin{array}{c}
d\pm d_3  \\
\pm(d_1+\mi d_2)\end{array}\right).
\end{align}
Here $2m$ is the gap between the two eigen-energies at $\bk=\mathbf{0}$, and the Dirac fermions are time reversal invariant only when $d_3=m=0$. It appears that $|u_-\rangle$ may be singular as $k\rightarrow 0$. However, this is an artifact because by parameterizing $\mathbf{d}(\bk)$ as $d_1=d(\bk)\sin\theta_\bk\cos\phi_\bk$, $d_2=d(\bk)\sin\theta_\bk\sin\phi_\bk$, and $d_3=d(\bk)\cos\theta_\bk$, Eq.~(\ref{EL}) implies
\begin{align}
|u_+\rangle=\left(\begin{array}{c}\cos\frac{\theta_\bk}{2}\\ \sin\frac{\theta_\bk}{2}\me^{\mi\phi_\bk}
 \end{array}\right),\quad
|u_-\rangle=\left(\begin{array}{c}
\sin\frac{\theta_\bk}{2}\\ -\cos\frac{\theta_\bk}{2}\me^{\mi\phi_\bk}
 \end{array}\right),
\end{align}
both of which are then well-behaved.

In this model, we choose the 2D momentum $\bk=(k_x,k_y)^T$ as the parameters to evaluate the Sj$\ddot{\text{o}}$qvist QGT.
It is straightforward to show that $g^\text{S}_{ij}=g^\text{FR}_{ij}+g^\text{FS}_{ij}-\mi\Omega_{ij}$, where
\begin{align}
g^\text{FR}_{ij}=&\frac{\beta^2}{4d^2}\text{sech}^2(\beta d)d_id_j,\notag\\
g^\text{FS}_{11}=&\frac{d^2_1d^2_3+d^2d^2_2}{4d^4(d^2-d^2_3)},\notag\\
g^\text{FS}_{22}=&\frac{d^2_2d_3^2+d^2d^2_1}{4d^4(d^2-d^2_3)},\notag\\
g^\text{FS}_{12}=&g^\text{FS}_{21}=\frac{d_1d_2}{4d^2},\notag\\
\Omega_{12}=&-\Omega_{21}=\tanh(\beta d)\frac{d_3}{4d^3}.
\end{align}
The contributions from the Fubini-Study metric in this case are also temperature-independent. However, the Sj$\ddot{\text{o}}$qvist QGT now has a non-zero imaginary part if $m\neq 0$. At first look,  $g^\text{FS}_{11,22}$ seems singular as $k\rightarrow 0$. However, this is resolved by noting that
\begin{align}
\lim_{k=0}g^\text{FS}_{11}=\lim_{k=0}g^\text{FS}_{22}=\frac{d^2(d^2_1+d^2_2)}{4d^4(d^2-d^2_3)}=\frac{1}{4d^2},
\end{align}
where $d^2=d^2_1+d^2_2+d^2_3$ has been applied. Thus, the Sj\"oqvist QGT is non-singular for gapped systems when $k\rightarrow 0$.

\begin{figure}[t]
\centering
\includegraphics[width=3.2in]{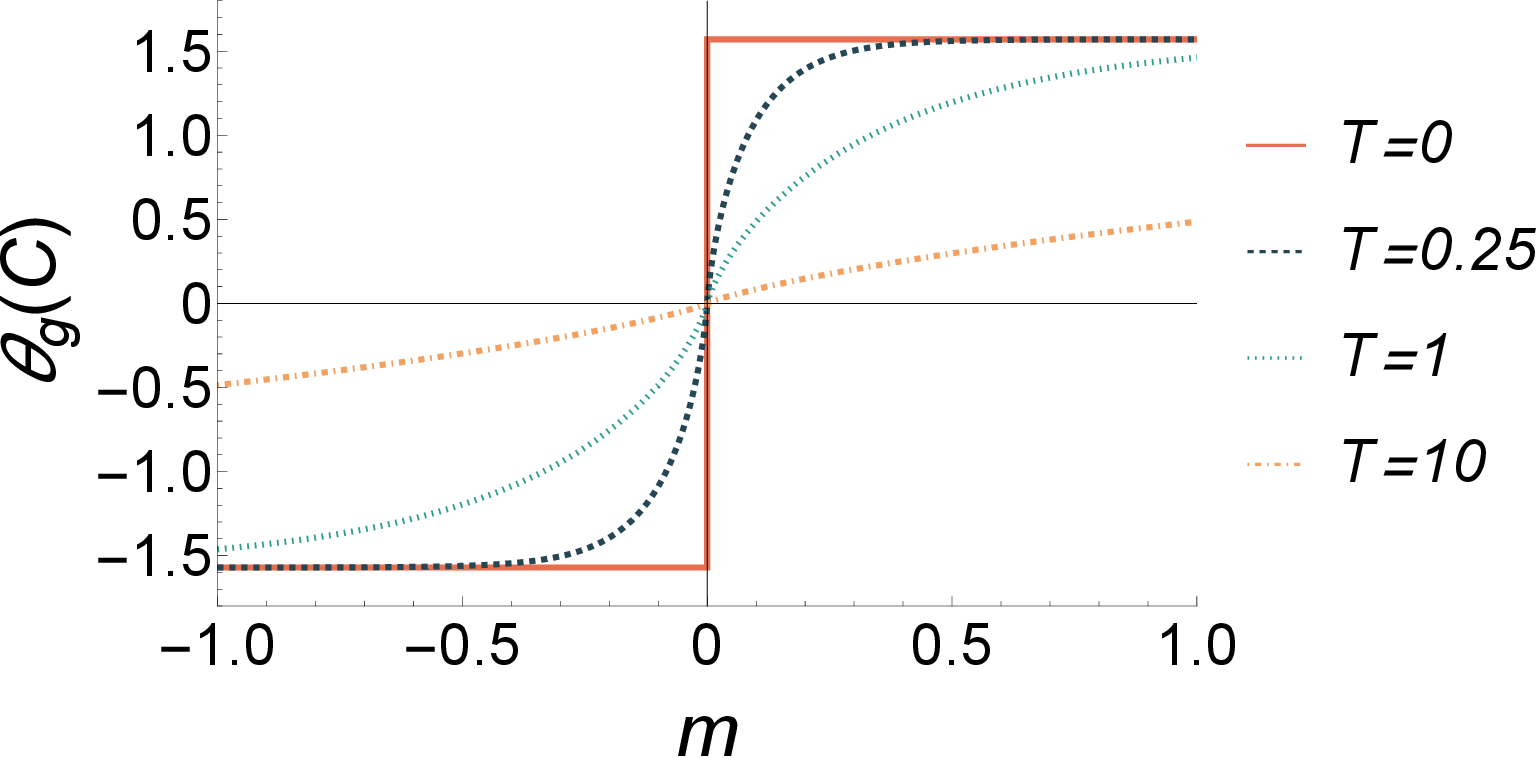}
\caption{$\theta_g$ (defined in Eq.~\eqref{aofSm}) of the 2D model as a function of $m$ for selected temperatures.
}
\label{Fig2b}
\end{figure}

We choose $g^\text{S}_{11}$ as a representative of the Sj\"oqvist QGT and show its contour plot at $(k_x,k_y)=(1.0,0.3)$ in Fig.~\ref{Fig2a}. It is symmetric about the time-reversal-invariant line $m=0$, and its dependence on $T$ is similar to that of the 1D SSH model. As $T\rightarrow 0$, it approaches $g^\text{FS}_{11}$ since $g^\text{FR}_{11}\rightarrow 0$. There is a peak around $T\in [1.0,1.4]$, and $g^\text{FS}_{11}\rightarrow 0$ as $T\rightarrow \infty$. $g^\text{S}_{11}$ is also insensitive to the topological properties of the system. We found that $g^S_{12}$ of this case does not exhibit additional features.

Fig.~\ref{Fig2b} shows $\theta_g(C)$ from Eq.~\eqref{aofSm} of the 2D model as a function of $m$, where $C$ is the loop formed by points at infinity.
 As $T\rightarrow 0$, $\theta_g(C)$ approaches half the Berry phase of the ground state. Explicitly, Eq.~(\ref{aofSm}) yields
\begin{align}\label{aofSm2}
\theta_g(C)=&\pi\int_0^\infty\dif kk\tanh(\beta d)\frac{m}{(k^2+m^2)^{\frac{3}{2}}}\notag\\\rightarrow&\frac{\pi}{2}\text{sgn}(m)\notag\\
=&\frac{1}{2}\theta^-_\text{B}(C),
\end{align}
When $m=0$, the energy gap closes at $\mathbf{k}=\mathbf{0}$, and the two energy eigenvalues join together there. Thus, the manifold of the spectrum of the Dirac fermion no longer has a consistent orientation. As a consequence, the Berry phase is no longer well-defined. The red solid line ($T=0$) in Fig.~\ref{Fig2b} shows that $\theta_g$ jumps by a factor of $\pi$ when crossing the $m=0$ line due to the geometric phase transition.
At finite temperatures, $\theta_g(C)$ shows some resemblance of $\theta^-_\text{B}(C)$ but changes smoothly as $m$ varies from $m<0$ to $m>0$.

\begin{figure*}[t]
\centering
\includegraphics[width=2.3in]{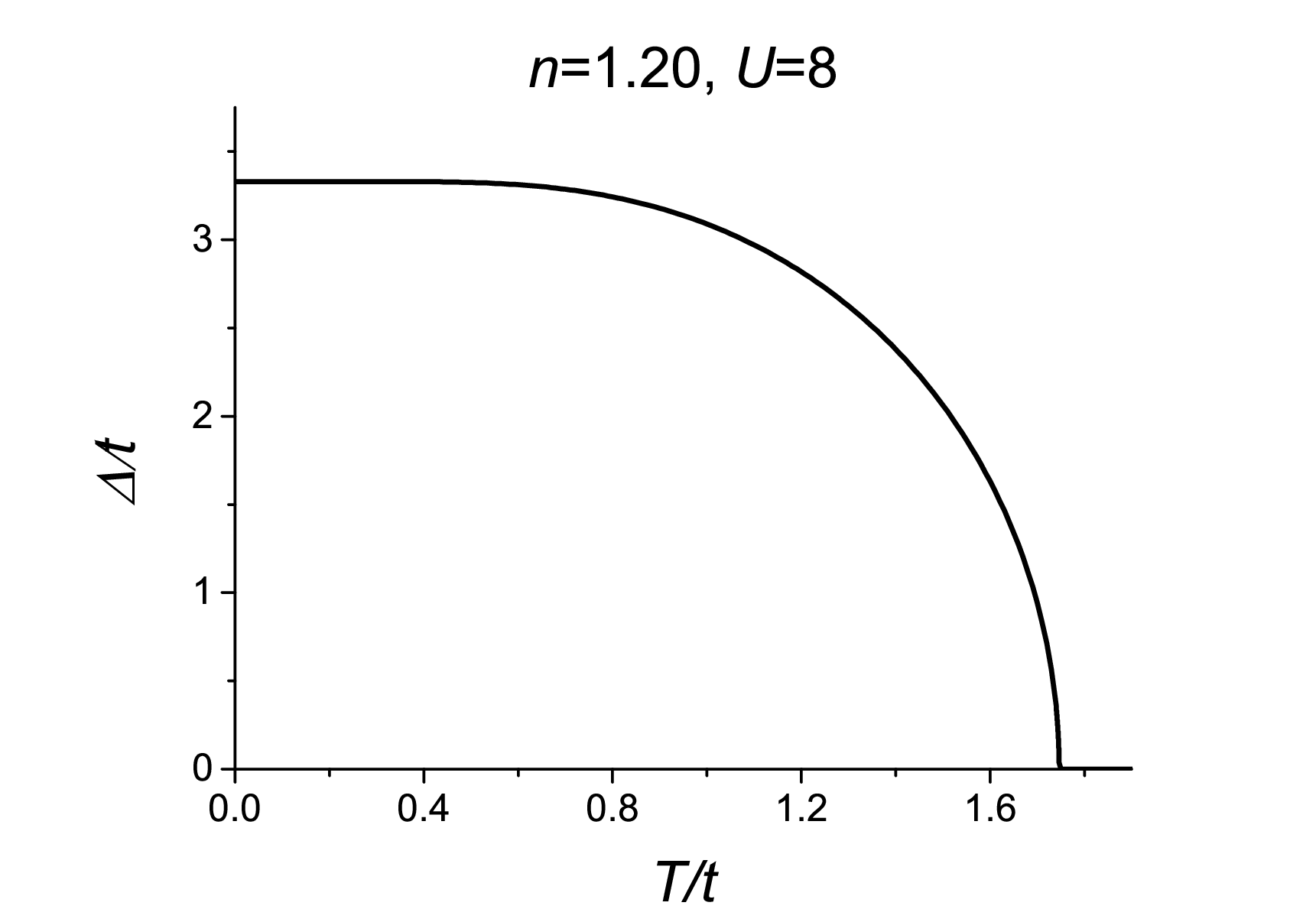}\includegraphics[width=2.3in]{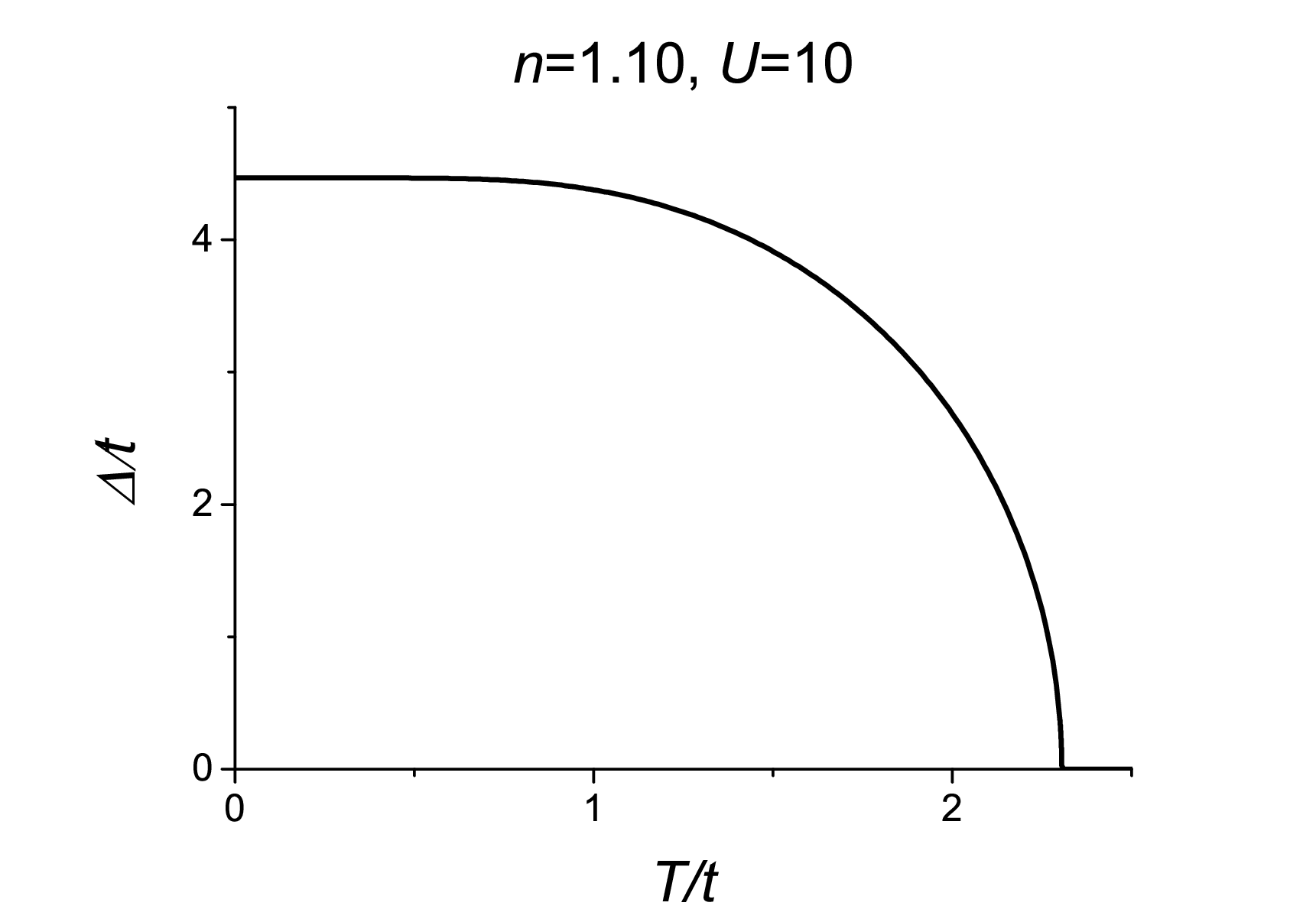}\includegraphics[width=2.3in]{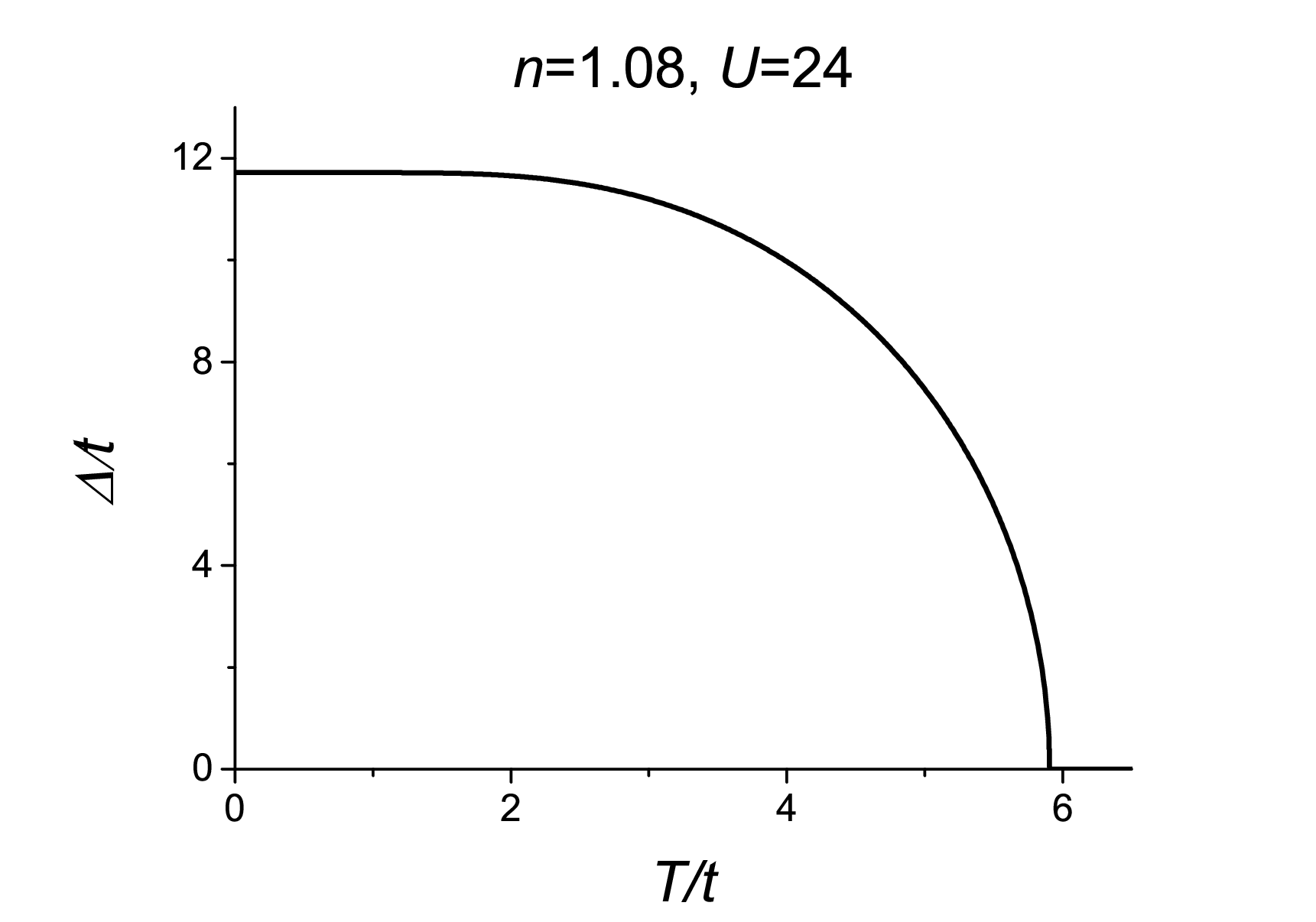}\\
\includegraphics[width=2.3in]{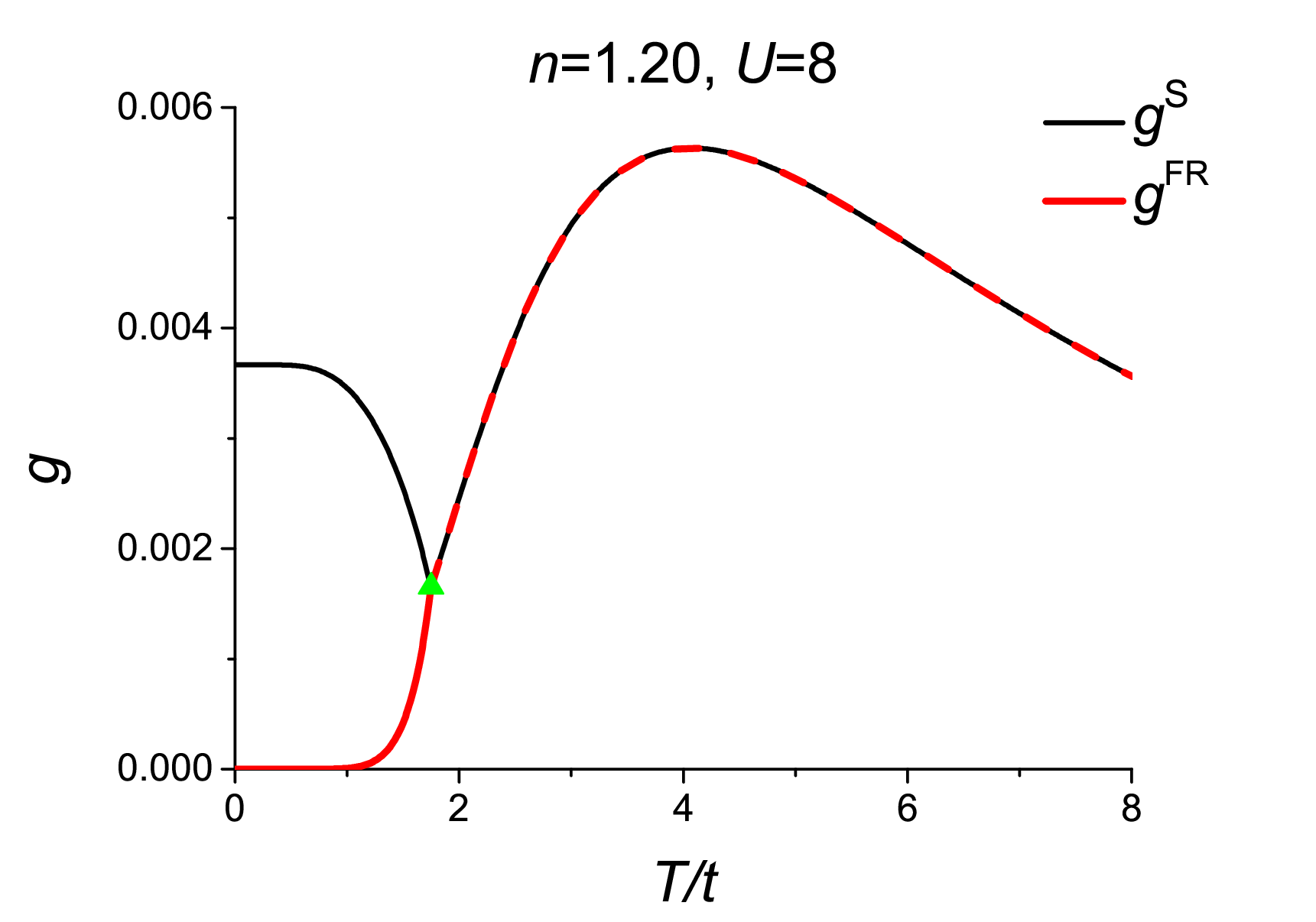}\includegraphics[width=2.3in]{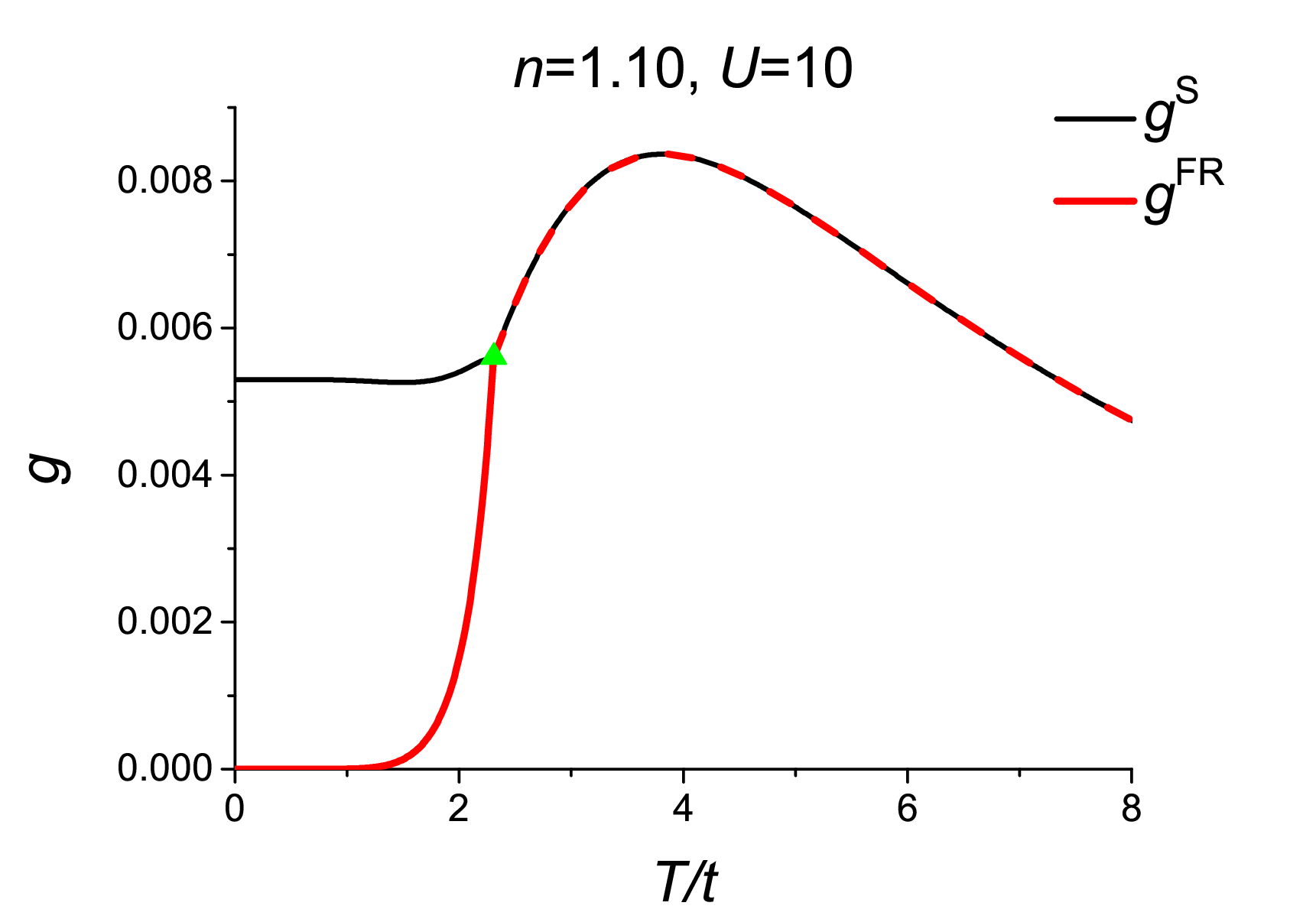}\includegraphics[width=2.3in]{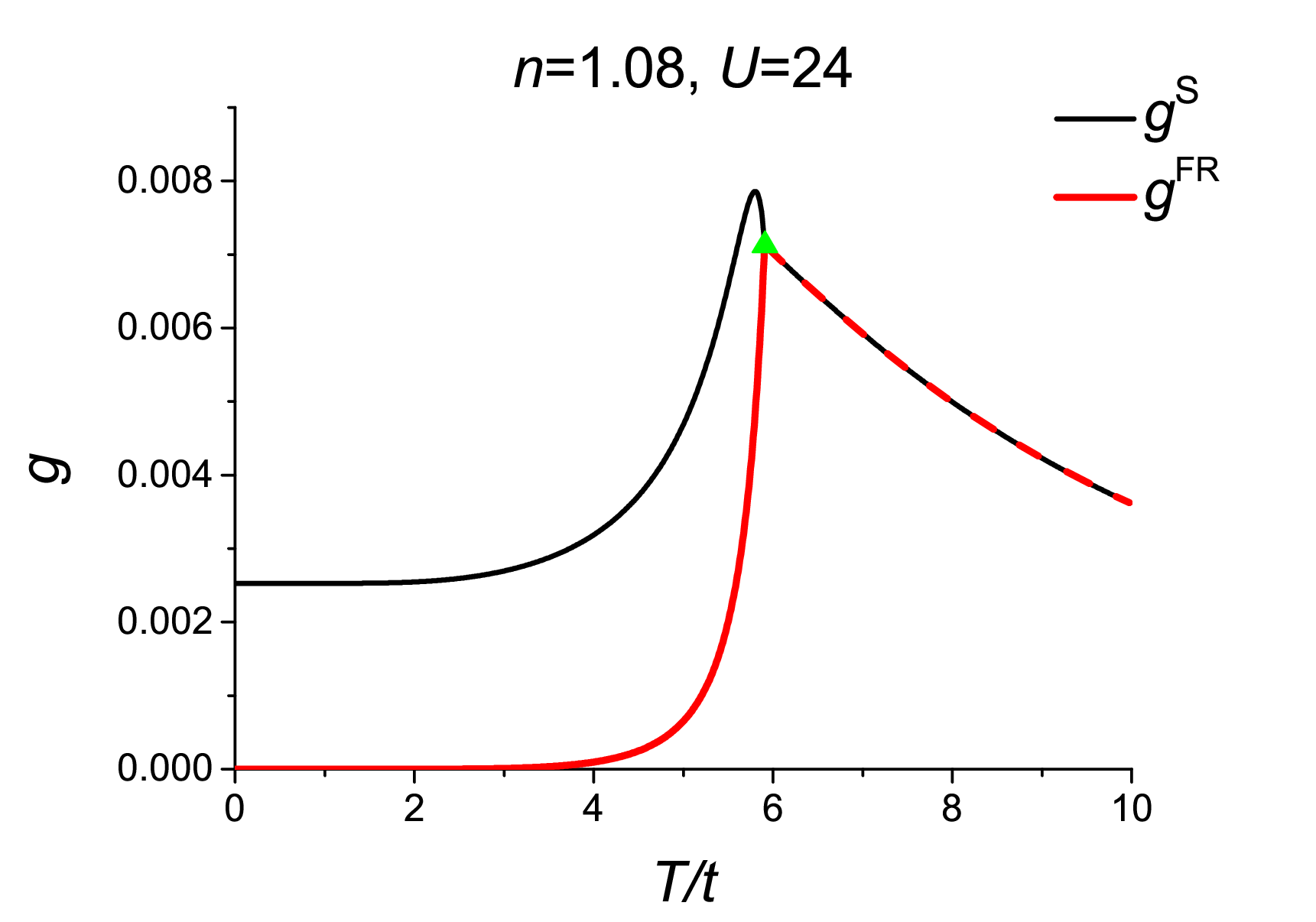}
\caption{Top row: The order parameter of the 3D superconductor model as a function of $T$, where the pairing strength increases from left to right ($U/t$=8, 10, 24) and the number density is nearly constant ($n=1.20$, $1.10$ and $1.08$). Bottom row: The corresponding $g^\text{S}$ and $g^\text{FR}$ vs. $T$ at $k_x=k_y=k_z=\frac{\pi}{4}$. Here the green dots label the critical temperature $T_c$. The black and red lines respectively represent $g^\text{S}$ and $g^\text{FR}$. Above $T_c$, $g^\text{S}=g^\text{FR}$.}
\label{Fig2}
\end{figure*}

\subsection{3D Superconductor}
Finally, we consider a 3D spin-singlet Fermi superfluid with the mean-field Hamiltonian $\hat{H}=\sum_\bk\Psi^\dag_\bk H(\mathbf{k})\Psi_\bk$, where $\Psi_\bk=(\psi_{\bk\uparrow},\psi^\dag_{-\bk\downarrow})^T$is the Nambu spinor of two-component fermions with spins $\uparrow$ and $\downarrow$, and
\begin{equation}\label{SCH1}
H(\mathbf{k})=\mathbf{d}(\mathbf{k}) \cdot \boldsymbol{\sigma}=d_{1}(\mathbf{k}) \sigma_{1}+d_{3}(\mathbf{k}) \sigma_{3}.
\end{equation}
Here $d_{1}=\Delta$ is the order parameter or gap function, which depends on temperature, and $d_{3}=\epsilon_{\mathbf{k}}=-2t(\cos{k_{x}}+\cos{k_{y}}+\cos{k_{z}})-\mu$, where $t$ is the nearest-neighbor hopping coefficient, and $\mu$ is the chemical potential.
The gap function is determined by the gap equation  $\Delta=U\sum_\bk\langle\psi_{\bk\uparrow}\psi_{-\bk\downarrow}\rangle$, where $U$ is the pairing coupling constant. The energy dispersion and the corresponding energy levels are respectively given by
$E_{\pm}(\mathbf{k})=\pm d(\mathbf{k})=\pm\sqrt{\Delta^{2}+\epsilon_{\mathbf{k}}^{2}}$, and
\begin{align}\label{4}
&|u_{\pm}\rangle=\frac{1}{\sqrt{2 d\left(d\pm d_{3}\right)}}\left(\begin{array}{c}
d\pm d_{3} \\
\pm d_{1}
\end{array}\right).
\end{align}
Here the dependence on $\mathbf{k}$ of Eq.~(\ref{4}) has been suppressed for simplicity.

In terms of the Nambu spinor, the model is equivalent to a two-band system. The associated thermal state is described by the density matrix
$\rho_\mathbf{k}=\frac{1}{2}\left[1-\tanh\left(\beta d(\mathbf{k})\right)\hat{\mathbf{d}}(\mathbf{k}) \cdot \boldsymbol{\sigma}\right]$,
whose eigenvalues are $\lambda_{\mathbf{k}\pm}=\frac{1}{2}\left[1\mp\tanh(\beta d(\mathbf{k}))\right]$. At temperature $T$ with a given $t$, $\Delta$, and $\mu$ can be obtained by solving the number equation
\begin{equation}
n=\sum_{\bk}\left[1-\frac{\epsilon_{\bk}}{d(\bk)}\big(1-2f(d(\bk))\big)\right]
\end{equation}
and the gap equation
\begin{equation}
\frac{1}{U}+\sum_{\bk}\frac{1-2f(d(\bk))}{2d(\bk)}=0
\end{equation}
simultaneously.
Here $n$ is the number density, and $f(x)=(\me^{\frac{x}{T}}+1)^{-1}$ is the Fermi distribution function. $T_c$ is the superconducting transition temperature determined by $\Delta>0$ if $T<T_c$ and $\Delta=0$ if $T>T_c$. In this model, we are interested in the behavior of the Sj$\ddot{\text{o}}$qvist QGT across $T_c$.

Taking $\mathbf{k}$ as the parameter, the Sj$\ddot{\text{o}}$qvist QGT is evaluated according to Eq.~(\ref{Sm1}). Although the Sj$\ddot{\text{o}}$qvist QGT is found to be real-valued in this 3D model, complex-valued Sj$\ddot{\text{o}}$qvist QGT may be possible in more complicated systems. Moreover,  $g^\text{S}_{ij}=g^\text{FR}_{ij}+g^\text{FS}_{ij}$
for $i,j=x,y,z$,
where
\begin{align}\label{SmE2b}
g^\text{FR}_{ij}=&\sin{k_{i}}\sin k_j\frac{t^{2}\beta^{2}d_{3}^{2}\text{sech}^{2}(\beta d)}{d^{2}},\notag\\
g^\text{FS}_{ij}=&\sin{k_{i}}\sin k_jt^2\frac{d^2_1}{d^4}=\frac{\sin{k_{i}}\sin k_j\Delta^2t^2}{(\Delta^2+\epsilon^2_\mathbf{k})^2}.
\end{align}
Interestingly, $g^\text{FS}_{ij}$ is proportional to $\Delta^2$ in this model and is expected to reflect the change of the order parameter. On the other hand, $g^\text{FR}_{ij}$ only has implicit dependence on $\Delta$ through the energy dispersion in the thermal factor. Combining those effects,  Sj$\ddot{\text{o}}$qvist QGT should exhibit different behavior as the system crosses $T_c$.

Figure \ref{Fig2} shows $g^\text{S}_{ij}$ and $g^\text{FR}_{ij}$of three selected sets of parameters of the model as functions of temperature. Moreover, we show the temperature dependence of $\Delta$ in the top row of Figure \ref{Fig2}. In the three cases, the number densities are basically the same. The pairing coupling constant is set to $U/t=8$, $10$, and $24$ to represent relatively weak to strong pairing effects. From the temperature at which $\Delta$ vanishes, the critical temperatures are extracted as $T_c/t=1.75$, $2.31$, and $5.90$ for the three cases. At $T=0$, the order parameters are respectively $\Delta/t=3.33$, $4.47$, and $11.72$. For $k_x=k_y=k_z=\frac{\pi}{4}$, all nine components of the metrics are equal according to Eq.~(\ref{SmE2b}).
As mentioned before, $g^\text{FS}$ is proportional to $\Delta^2$, which results in different behavior across $T_c$ for the three cases. One can see that $g^\text{FS}$ vanishes above $T_c$ since $\Delta=0$, so $g^\text{S}=g^\text{FR}$ above $T_c$.
At low temperatures, $g^\text{FR}\rightarrow 0$ for the same reason as explained previously. Therefore, $g^\text{S}=g^\text{Fs}$ in the zero-temperature limit. Near $T_c$, the contributions from $g^\text{FS}$ and $g^\text{FR}$ experience significant decreases and increases, respectively. In the relatively weak-pairing regime (for example, $U/t=8$ with $T_c=1.75t$), the combination of these two opposite effects results in a valley near $T_c$. This is due to the factor $\frac{\text{sech}^2(d/T)}{d^2}$ in the expression of $g^\text{FR}$, which is a monotonically increasing function near $T_c$. Its effect becomes more dominant with larger $T_c$ when $U$ increases. Consequently, in the relative medium-coupling regime (for example, $U/t=10$ with $T_c=2.31t$), the valley of $g^\text{S}$ disappears. Finally, in the relative strong-coupling scenario (for example, $U/t=24$ with $T_c=5.90t$), a peak emerges below $T_c$. The contrasts among the three cases demonstrate the rich behavior of the Sj$\ddot{\text{o}}$qvist QGT as the system crosses the superfluid transition point with different pairing strengths.

\section{Implications}\label{sec:implications}
The pure-state QGT has been experimentally measured in various physical platforms, as mentioned in the Introduction. The mixed-state QGT is a developing concept and needs more research explorations in the future. The Sj\"oqvist distance has been proposed to be related to some physical observables, such as the maximal probability to find particles in a Mach-Zehnder interferometer \cite{PhysRevResearch.2.013344} and the magnetic susceptibility and fidelity susceptibility \cite{PhysRevResearch.2.013344,PhysRevE.76.022101}. The Sj\"oqvist QGT derived here inherits the Sj\"oqvist metric as its real part and broadens its geometric implications in realistic quantum systems. The imaginary part of the Sj\"oqvist QGT, on the other hand, is related to the thermal Berry phase in certain situations and may be inferred from the Berry phases of individual spectrum levels and their thermal distribution. The decomposition of the Sj\"oqvist metric into the Fisher-Rao metric and the Fubini-Study metric further helps categorize the contributions from the variations of the states and distributions, which serves as useful information for designing future quantum systems with robust features against parameter variations.

The rich physics of mixed quantum states allows multiple QGTs from different gauge transformations and parallel conditions. The $U^N(1)$-invariant Sj\"oqvist QGT is distinct from the $U(N)$-invariant Uhlmann QGT~\cite{OurQGT23}. One has to verify the conditions when comparing different QGTs with experimental data, as experiments may impose particular constraints on the quantum processes. While the real parts of both Sj\"oqvist and Uhlmann QGTs approach the Fubini-Study metric in the zero-temperature limit, Ref.~\cite{OurQGT23} shows that the Sj\"oqvist distance cannot exceed the Bures distance from the Uhlmann QGT.
On the other hand,
only the imaginary part of the Sj\"oqvist QGT approaches the Berry curvature as $T\rightarrow 0$, which is the imaginary part of the pure-state QGT, because the imaginary part of the Uhlmann QGT vanishes in general systems.

\section{Conclusion}\label{sec:conclusion}
Through the formalism of the Sj\"oqvist distance, we extend the concept of the QGT from pure states to mixed states at finite temperatures and construct a U$^N(1)$-invariant QGT applicable to thermal equilibrium states. Based on its geometric structure, a Pythagorean-like equation connecting different types of distances is presented. The real part of the QGT contains the contributions from the Fisher-Rao
metric and the Fubini-Study metric from each energy level, while the imaginary part defines a gauge-invariant quantity from the weighted summation of the Berry curvatures. Our examples illustrate the temperature-dependence of the metrics and geometric phase associated with the QGT. Furthermore, the Sj\"oqvist QGT is expected to serve as a tool for discovering and quantifying geometric information of mixed states.

\section{Acknowledgments}
H.G. was supported by the National Natural Science
Foundation of China (Grant No. 12074064) and the Innovation Program for Quantum
Science and Technology (Grant No. 2021ZD0301904). X. Y. H. was supported by the Jiangsu Funding Program for Excellent Postdoctoral Talent (Grant No. 2023ZB611). C.C.C. was supported
by the National Science Foundation under Grant No.
PHY-2310656.

\appendix
\section{Some details of the Sj$\ddot{\text{o}}$qvist distance}\label{appa}
By expanding the right-hand-side of Eq.~(\ref{Sdis}), we get
 \begin{widetext}
 \begin{align}\label{Sdis1a}
&2-\sqrt{\lambda_n(t)\lambda_n(t+\dif t)}\left[\me^{\mi\dot{\theta}_n(t)\dif t}\langle n(t)|n(t+\dif t)\rangle+\me^{-\mi\dot{\theta}_n(t)\dif t}\langle n(t+\dif t)|n(t)\rangle\right]\notag\\
=&2-\sqrt{\lambda_n(t)\lambda_n(t+\dif t)}\big|\langle n(t)|n(t+\dif t)\rangle\big|\left[\me^{\mi\dot{\theta}_n(t)\dif t}\me^{\mi\arg \langle n(t)|n(t+\dif t)\rangle}+\me^{-\mi\dot{\theta}_n(t)\dif t}\me^{-\mi\arg \langle n(t)|n(t+\dif t)\rangle}\right]\notag\\
=&2-2\sqrt{\lambda_n(t)\lambda_n(t+\dif t)}\big|\langle n(t)|n(t+\dif t)\rangle\big|\cos\left[\dot{\theta}_n(t)\dif t+\arg \langle n(t)|n(t+\dif t)\rangle\right].
\end{align}
\end{widetext}
The infimum of the left-hand-side of Eq.~(\ref{Sdis}) is obtained when $\dot{\theta}_n(t)\dif t+\arg \langle n(t)|n(t+\dif t)\rangle=0$. Since
 \begin{align}
 \arg \langle n(t)|n(t+\dif t)\rangle&=\arg(1+ \langle n(t)|\dot{n}(t)\rangle\dif t+O(\dif t^2))\notag\\
  &\approx \arg\me^{\mi(-\mi\langle n(t)|\dot{n}(t)\rangle\dif t)}\notag\\
  &=-\mi\langle n(t)|\dot{n}(t)\rangle\dif t,
\end{align}
the minimization condition is equivalent to
\begin{align}\label{Sdispcb}
\mi\dot{\theta}(t)+\langle n(t)|\dot{n}(t)\rangle=0, \quad \text{for} \quad n=0,\cdots, N-1,
\end{align}
which is precisely the parallel-transport condition associated with each individual pure state in the ensemble.

\section{Another construction of the $U^N(1)$-invariant QGT}\label{appb}
Similar to Eq.~(\ref{Sdis3c}),
for $W=\sum_n\sqrt{\lambda_n}|n\rangle\langle n|$, the raw metric is given by
\begin{align}\label{Sdisrm}
g_{\mu\nu}=\sum_n\left[\frac{\partial_\mu\lambda_n\partial_\nu\lambda_n}{4\lambda_n}+\lambda_n\langle \partial_\mu n|\partial_\nu n\rangle\right].
\end{align}
Under the U$^N(1)$ gauge transformation $W\rightarrow W'=W\mathcal{U}$ with  $\mathcal{U}=\text{diag}(\text{e}^{\text{i}\chi_0},\text{e}^{\text{i}\chi_1},\cdots,\text{e}^{\text{i}\chi_{N-1}})$,
the first term of $g_{\mu\nu}$, the Fisher-Rao metric, is already invariant while the second term is not. However, it is not hard to see that the imaginary part of the second term, which is $\Omega_{\mu\nu}$, is also invariant under the transformation.
To impose a proper modification, we first note that the raw distance changes as
\begin{align}\label{Sdisr13}
&\dif^2(W+\dif W,W)\rightarrow \dif^{\prime 2}(W+\dif W,W)\notag\\
=&\sum_n\big[\partial_\mu \sqrt{\lambda_n}\partial_\nu \sqrt{\lambda_n}+\lambda_n(\langle\partial_\mu n|\partial_\nu n\rangle+\partial_\mu \chi_n\partial_\nu\chi_n \notag\\-&\mi\omega_{n\mu}\partial_\nu\chi_n-\mi\omega_{n\nu}\partial_\mu\chi_n) \big]\dif R^\mu \dif R^\nu.
\end{align}
Similar to the pure-state case \cite{QGT10}, to maintain the gauge-invariance, we can modify the raw metric as
\begin{align}
\gamma_{\mu\nu}=g_{\mu\nu}+\sum_n\lambda_n\omega_{n\mu}\omega_{n\nu}
\end{align}
Comparing with Eq.~(\ref{Sm1}), one can see that $\gamma_{\mu\nu}$ is nothing but the Sj$\ddot{\text{o}}$qvist metric.  The gauge-invariance becomes clear by noting that the extra terms of $g_{\mu\nu}$ in Eq.~(\ref{Sdisr13}) are cancelled by the changes of the Berry connections. For the $n$th spectral level,  $\omega_{n\mu}\rightarrow \omega'_{n\mu}=\omega_{n\mu}+\mi\partial_\mu \chi_n$, then $\gamma_{\mu\nu}=g^\text{S}_{\mu\nu}$ is indeed invariant under the transformation $W'=W\mathcal{U}$.

%

\end{document}